\documentclass[11pt]{article}

\usepackage[margin=1in]{geometry}

\usepackage[utf8]{inputenc}
\usepackage[T1]{fontenc}
\usepackage{amsmath,amssymb}
\usepackage{graphicx}
\usepackage{booktabs}
\usepackage{multirow}
\usepackage{subcaption}
\usepackage{algorithm}
\usepackage{algpseudocode}
\usepackage{xcolor}
\usepackage{url}
\usepackage{authblk}

\usepackage{tikz}
\usetikzlibrary{positioning,fit,backgrounds}

\title{Query-Efficient Quantum Multi-Objective Optimization for
Materials Configuration Search via Bayesian Weight Selection}

\author{Onur Boyar}
\author{Seiji Takeda}
\author{Akihiro Kosugi}
\author{Indra Priyadarsini S}
\author{Lisa Hamada}
\author{Sina Klampt}
\affil{IBM Research}
\affil{\normalfont\small\texttt{Onur.Boyar@ibm.com}, \texttt{SEIJITKD@jp.ibm.com},
  \texttt{A1KOSUGI@jp.ibm.com},\\
  \texttt{indra.ipd@ibm.com}, \texttt{Lisa.Hamada@ibm.com}, \texttt{Sina.Klampt@ibm.com}}
\date{}

\newcommand{\bs}{\boldsymbol{\sigma}}
\newcommand{\bw}{\mathbf{w}}
\newcommand{\ket}[1]{\lvert #1 \rangle}

\makeatletter
\renewcommand\section{\@startsection{section}{1}{\z@}%
  {-2.0ex \@plus -0.5ex \@minus -0.2ex}%
  {0.8ex \@plus 0.2ex}%
  {\normalfont\large\bfseries}}
\renewcommand\subsection{\@startsection{subsection}{2}{\z@}%
  {-1.8ex \@plus -0.5ex \@minus -0.2ex}%
  {0.6ex \@plus 0.2ex}%
  {\normalfont\normalsize\bfseries}}
\renewcommand\paragraph{\@startsection{paragraph}{4}{\z@}%
  {1.2ex \@plus 0.4ex \@minus 0.2ex}%
  {-0.6em}%
  {\normalfont\normalsize\bfseries}}
\makeatother

\begin{document}
\maketitle

\begin{abstract}
Materials design is a multi-objective problem, and recent advances in quantum optimization make it
possible to bring quantum solvers to bear on such problems. The realistic near-term win, though, is
not raw quantum advantage but making an expensive, noisy quantum solver \emph{usable} by spending its
calls wisely. Quantum Approximate Multi-Objective Optimization (QAMOO)~\cite{qamoo} traces a Pareto front by
solving a family of scalarized Ising problems, one per objective-weight vector, with QAOA --- but
chooses those weights \emph{non-adaptively}, sampling them at random, blind to what it has already
found. We recast weight selection as \emph{multi-objective Bayesian optimization}: our method, \mbox{BW-QAMOO}
(Bayes-Weighted QAMOO), picks each next weight by \emph{expected hypervolume improvement} (EHVI) under a
Gaussian-process (GP) surrogate over the weight-to-objective map. On a frustrated
alloy-configuration benchmark across $m\in\{3,4\}$ objectives and $n\in\{10,11,12\}$ sites, BW-QAMOO
reaches the same Pareto hypervolume as QAMOO's random sampling with a \emph{median $33\%$ fewer}
scalarized solves across the $90/95\%$ targets (and $60$--$74\%$ fewer than a uniform-grid baseline), on
a genuine (simulated) QAOA solver; a paired Wilcoxon test confirms the saving is significant, and
the advantage grows with the objective count $m$ as the weight simplex expands and blind sampling
wastes more of the budget. This is a \emph{query-efficiency} result, and we
close with the open challenges on the path to real alloy data and hardware.
\end{abstract}

\section{Introduction}

\begin{figure}[t]
\centering
\includegraphics[width=\textwidth]{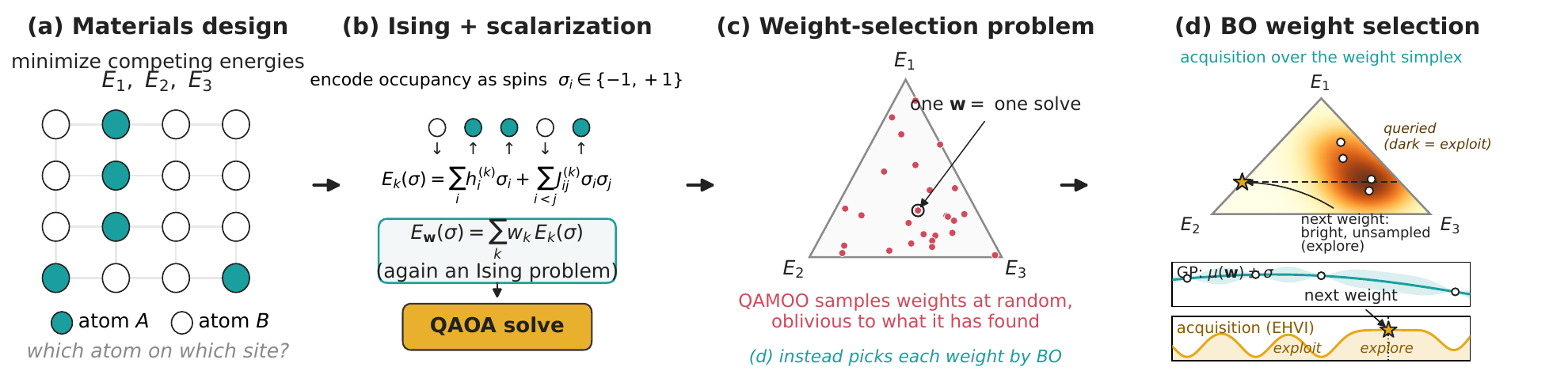}
\caption{The QAMOO pipeline and where BW-QAMOO's Bayesian weight selection enters.
\textbf{(a)} A materials-design instance under several competing configurational energies.
\textbf{(b)} Occupancy is encoded as Ising spins $\sigma_i\in\{-1,+1\}$; a weight vector $\bw$
scalarizes the $m$ energies into a single Ising cost, minimized by one QAOA run.
\textbf{(c)} QAMOO chooses the weights non-adaptively, sampling them at random (red).
\textbf{(d)} BW-QAMOO instead fits a Gaussian-process surrogate over the weight simplex and picks the
next weight by an expected-hypervolume-improvement acquisition (bright regions indicate high
acquisition values), so the query goes to
a \emph{bright, still-unsampled} region (explore) rather than the well-sampled cluster (dark, exploit).}
\label{fig:pipeline}
\end{figure}

Designing a material means navigating trade-offs, not optimizing a single number. An alloy is
chosen for a compromise among phase stability, cost, density, and chemical disorder, and the
design variable is combinatorial: which element sits on which lattice site. This
configuration-search problem maps cleanly onto an Ising model, $E(\bs)=\sum_i h_i\sigma_i+\sum_{i<j}
J_{ij}\sigma_i\sigma_j$, with spins $\sigma_i\in\{-1,+1\}$ encoding site occupancy and couplings
$J_{ij}$ the effective interactions~\cite{lucas}. This is exactly the class of problem quantum
optimizers are built to attack, which is what has made materials design one of the most active
near-term targets for quantum optimization (Appendix~\ref{app:ising} reviews the Ising model and this
mapping for the unfamiliar reader).

\paragraph{Quantum optimization for materials, and why it needs help.}
The evidence for that interest is concrete. Quantum annealers have searched crystal elemental
configurations~\cite{crystalconfig} and optimized high-entropy-alloy lattices~\cite{heaanneal}, and
gate-model devices attack the same energies with the Quantum Approximate Optimization Algorithm
(QAOA)~\cite{qaoa}. Two facts temper this promise, however. First, on today's noisy hardware these
methods rarely beat good classical solvers outright, so raw quantum advantage is not yet the right
thing to claim. Second, and more practically, \emph{each} solve is expensive: a single QAOA run
couples an iterative classical parameter search to repeated circuit sampling and, on real devices,
to cloud-QPU queue latency, which together cost seconds to minutes and real dollars per
solve~\cite{qaoa}. Taken together, these two facts point to a different near-term payoff. The goal
is not to outrun classical solvers but to make the quantum tool \emph{usable}, by wrapping it in
classical AI that spends its scarce, costly calls wisely. We therefore treat the quantum solver as
an expensive oracle and let classical learning decide what to ask it.

\paragraph{The multi-objective bottleneck, and QAMOO.}
Deciding what to ask is harder than it first appears, because a material is rarely chosen to minimize
one quantity. The same configuration must do well on several competing energies at once, and
improving one typically worsens another. This is a poor fit for the bulk of quantum optimizers,
which are built for a \emph{single} objective. The right target is therefore not a single minimizer
but the \emph{Pareto front}: the set of configurations for which no objective can improve without
sacrificing another. Quantum Approximate Multi-Objective Optimization (QAMOO)~\cite{qamoo} traces
this front by \emph{scalarization}. A weight vector $\bw$ on the simplex collapses the $m$ objectives
into a single Ising cost $E_{\bw}=\sum_k w_k E_k$; QAOA minimizes it; and the solution is recorded as
one Pareto point (Figure~\ref{fig:pipeline}). Sweeping $\bw$ over the simplex then produces the front.
Our benchmark makes this concrete with $m\in\{3,4\}$ competing energies, a deliberately simplified
stand-in for the objectives a crystal or alloy designer trades off (formation energy, elastic
mismatch, chemical short-range order, and cost), each expressed as its own Ising energy
(Appendix~\ref{app:objectives}).

The catch is that each scalarized solve is a separate costly quantum query, and QAMOO chooses the
weights \emph{non-adaptively}. Contrary to a natural reading, QAMOO does not grid the simplex; it
\emph{samples $\bw$ at random}, ``uniformly at random from the feasible set,'' without enumerating an
exponential grid~\cite{qamoo}. Either way the choice is oblivious to what has already been
discovered. This oversight grows more costly as the problem scales: for $m$ objectives the simplex is
$(m{-}1)$-dimensional, so blind sampling wastes an increasing share of the budget as $m$ grows.
QAMOO itself names better weight selection, ``non-uniform sampling of weights or other scalarization
strategies,'' as an explicit open direction~\cite{qamoo}, and it is exactly this gap we set out to
fill. We ask a simple question: \emph{can we reach the same front with fewer quantum queries by
choosing the weights adaptively?}

Our answer is yes, with a precise scope. We first \emph{implement} QAMOO on a materials-design
instance, then make it query-efficient through a novel integration with multi-objective
Bayesian optimization (BO). We treat weight selection as a sequential decision problem and choose
each next weight by an expected-hypervolume-improvement acquisition (EHVI~\cite{ehvi}) under a
Gaussian-process surrogate of the weight-to-objective map. We call the resulting method BW-QAMOO
(Bayes-Weighted QAMOO). Across $m\in\{3,4\}$ and $n\in\{10,11,12\}$, BW-QAMOO reaches the $90/95\%$
hypervolume targets with a median $33\%$ fewer solves than QAMOO (random), and $61$--$74\%$ fewer than
QAMOO (uniform grid) at $m{=}4$, on a genuine (simulated) QAOA solver, and the advantage grows with the
objective count $m$.

\paragraph{From alloy configuration to a (multi-objective) Ising problem.}
Encode each of $n$ lattice sites' binary occupancy by a spin $\sigma_i\in\{-1,+1\}$. In the
cluster-expansion picture, each objective's configurational energy is a sum of on-site preferences
and pairwise interactions --- an \emph{Ising Hamiltonian} --- and a weight vector $\bw$ on the simplex
($w_k\ge0,\ \sum_k w_k=1$) scalarizes the $m$ per-objective energies $E_k$ into a single cost:
\begin{equation}
E_k(\bs)=\sum_i h^{(k)}_i\,\sigma_i+\sum_{i<j}J^{(k)}_{ij}\,\sigma_i\sigma_j,
\qquad
E_{\bw}(\bs)=\sum_k w_k\,E_k(\bs)
=\sum_i\bar h_i\,\sigma_i+\sum_{i<j}\bar J_{ij}\,\sigma_i\sigma_j,
\label{eq:ising}
\end{equation}
the $\bw$-weighted sums of the per-objective fields and couplings (the effective cluster
interactions fitted from first principles in practice~\cite{lucas,crystalconfig}). Crucially, the
scalarized cost $E_{\bw}$ is \emph{again} of Ising form, so every scalarized subproblem is itself a
valid single-objective QAOA instance (each $\sigma_i$ maps to a qubit Pauli-$Z$, $E_{\bw}$ to a
diagonal cost Hamiltonian $H_C$; Appendix~\ref{app:qamoo}), and QAMOO~\cite{qamoo} traces the front by
solving $\arg\min_{\bs} E_{\bw}$ for many $\bw$. The only quantity our method changes is \emph{how many,
and which,} weights are solved.

\paragraph{Related work.}
Choosing where to sample next to trace a Pareto front with few expensive evaluations is the subject
of multi-objective BO. \emph{Scalarization-based} methods such as
ParEGO~\cite{parego} draw a weight each iteration and fit a surrogate to the scalarized objective ---
QAMOO's random weight sampling is exactly this draw step, and our loop replaces the random draw with
one chosen to maximize front progress; \emph{hypervolume-based} methods select by expected
hypervolume improvement (EHVI)~\cite{ehvi}, the standard multi-objective BO acquisition, now available in
differentiable parallel form~\cite{qehvi}, and connect to Pareto active learning and multi-objective
bandits~\cite{epal}. Our acquisition~\eqref{eq:ehvi} is EHVI over the weight simplex; our
contribution is to port this BO machinery to the \emph{weight-selection} layer of
QAMOO~\cite{qamoo}, where each ``evaluation'' is a full scalarized QAOA solve~\cite{qaoa}.

\section{BW-QAMOO: Bayesian-Optimized Weight Selection}
\label{sec:method}

BW-QAMOO casts weight selection as BO: it treats each scalarized QAOA
solve as an expensive black-box evaluation and, at each step, chooses the weight $\bw$ whose solve is
expected to enlarge the discovered Pareto front the most. Like any BO loop it has
two modular parts: a \emph{surrogate} that predicts a candidate's objective vector without spending a
query, and an \emph{acquisition} that scores each candidate's expected ``front progress'' under that
surrogate. Let $\mathcal{P}_t$ be the
objective vectors discovered after $t$ queries and $\mathrm{HV}(\cdot)$ the hypervolume relative to a
fixed reference point $\mathbf{r}$. The next weight maximizes the expected hypervolume improvement
(EHVI) --- the standard acquisition of multi-objective BO, and the reward signal
of a hypervolume bandit:
\begin{equation}
\bw_{t+1}=\arg\max_{\bw}\ \mathbb{E}\big[\mathrm{HV}(\mathcal{P}_t\cup\{\bs_{\bw}\})-\mathrm{HV}(\mathcal{P}_t)\big],
\label{eq:ehvi}
\end{equation}
where $\bs_{\bw}$ is the (unknown until queried) solution the solver returns for weight $\bw$.

\paragraph{Surrogate.} To score \eqref{eq:ehvi} without spending a query, BW-QAMOO keeps a surrogate
over $\bw\mapsto\bs_{\bw}$, refined online as real queries arrive: a \emph{Gaussian process},
one GP regressor per objective (Mat\'ern-$5/2$ kernel, an anisotropic length
scale per simplex coordinate), giving a posterior \emph{mean and variance} for each unqueried weight.
The acquisition scores each candidate by an \emph{optimistic} (upper-confidence-bound) estimate of its
hypervolume gain: it forms the optimistic objective vector $\hat{\bs}(\bw)=\boldsymbol{\mu}(\bw)-\kappa\,\boldsymbol{\sigma}(\bw)$
(per-objective posterior mean minus $\kappa$ standard deviations, $\kappa{=}1.5$; the objectives are
minimized, so subtracting is optimistic) and evaluates the hypervolume improvement that point would add,
$\mathrm{HV}(\mathcal{P}_t\cup\{\hat{\bs}(\bw)\})-\mathrm{HV}(\mathcal{P}_t)$. The $\kappa\boldsymbol{\sigma}$ term
supplies genuine exploration --- the selector can prefer a weight whose outcome is \emph{uncertain} over
one whose predicted mean gain is merely high --- realizing the explore/exploit trade-off of the EHVI
acquisition of multi-objective BO~\cite{ehvi,qehvi} in a cheap, deterministic (one hypervolume evaluation
per candidate) form, ported to the weight simplex (Appendix~\ref{app:gp}). Algorithm~\ref{alg:bandit}
summarizes the loop.

\paragraph{Continuous weight selection.} Because QAMOO samples weights \emph{continuously} from the
simplex, BW-QAMOO optimizes the acquisition over the continuous simplex too: each step draws a fresh
dense Dirichlet candidate pool, scores Eq.~\eqref{eq:ehvi}, and takes the argmax. This converges to the
continuous maximizer as the pool grows and keeps the comparison against QAMOO's random sampler on equal,
grid-free footing.

\begin{algorithm}[t]
\caption{BW-QAMOO: Bayesian-optimized weight selection}
\label{alg:bandit}
\begin{algorithmic}[1]
\State \textbf{Input:} budget $B$, solver $Q(\cdot)$, reference $\mathbf{r}$, pool size $P$
\State $\mathcal{S}\gets$ simplex corners $\{e_1,\dots,e_m\}$; query $\bs_{\bw}\gets Q(\bw)$ for each; $\mathcal{P}\gets\{\bs_{\bw}\}$
\While{$|\mathcal{S}|<B$}
  \State $h_{\text{cur}}\gets\mathrm{HV}(\mathcal{P};\mathbf{r})$; draw candidate pool $\mathcal{W}\sim\mathrm{Dirichlet}(\mathbf{1})^{\times P}$
  \For{each $\bw\in\mathcal{W}$}
     \State $\boldsymbol{\mu}(\bw),\boldsymbol{\sigma}(\bw)\gets$ GP posterior from $\mathcal{S}$; $\ \widehat{\bs}(\bw)\gets\boldsymbol{\mu}(\bw)-\kappa\,\boldsymbol{\sigma}(\bw)$ \Comment{optimistic (UCB) point, $\kappa{=}1.5$}
     \State $a(\bw)\gets\mathrm{HV}(\mathcal{P}\cup\{\widehat{\bs}(\bw)\};\mathbf{r})-h_{\text{cur}}$
  \EndFor
  \State $\bw^\star\gets\arg\max_{\bw}a(\bw)$;\quad $\bs^\star\gets Q(\bw^\star)$ \Comment{the one expensive query}
  \State $\mathcal{S}\gets\mathcal{S}\cup\{\bw^\star\}$;\quad $\mathcal{P}\gets\mathcal{P}\cup\{\bs^\star\}$
\EndWhile
\State \textbf{return} non-dominated subset of $\mathcal{P}$
\end{algorithmic}
\end{algorithm}

\section{Experimental setting}

\paragraph{A materials-configuration benchmark.} Our benchmark instantiates the multi-objective
alloy-configuration problem of Eq.~\eqref{eq:ising} on $n\in\{10,11,12\}$
lattice sites with $m\in\{3,4\}$ competing energies (a frustrated Ising model). Concretely, $E_1$ is a
configurational energy with random signed couplings $J_{ij}\in\{-1,+1\}\cdot\mathrm{U}(0.5,1.5)$ at
interaction density $0.5$ and site fields $h_i\sim\mathrm{U}(-0.3,0.3)$; the remaining $E_2,\dots,E_m$
are competing ``mixing-energy'' objectives with fixed distinct coupling matrices, standing in for
rival ordering tendencies. All are minimized; the goal is their Pareto front. This is a deliberate
\emph{synthetic} model of crystal design: each $E_k$ plays the role of a real design objective, mapped to
the same Ising form of fields and couplings but instantiated with \emph{random} couplings rather than
values fitted from first principles --- Appendix~\ref{app:objectives} gives the correspondence and is
explicit about what the idealization does and does not capture. We use random couplings so the exact
front is enumerable for scoring (Sec.~\ref{sec:results}); fitted cluster interactions are the natural
next step (Sec.~\ref{sec:disc}). Sweeping $m$ from $3$ to $4$ is deliberate --- the weight simplex
grows from two- to three-dimensional, exactly the growth non-adaptive sampling handles poorly.

\paragraph{Solver and strategies.} Each weight is solved by minimizing the scalarized Ising with a
genuine gate-based QAOA circuit on a shot-based state-vector simulator (Qiskit Aer~\cite{qiskit}); the
circuit construction, angle optimization, and read-out are detailed in Appendix~\ref{app:qamoo}. We compare three
weight-selection strategies, all sharing the same QAOA solver: (i) \emph{QAMOO (uniform grid)}, which walks a
granularity-$6$ uniform simplex grid in order (a naive non-adaptive reference; $28$ weights at $m{=}3$,
$84$ at $m{=}4$); (ii) \emph{QAMOO (random)}, i.i.d.\ Dirichlet sampling of the weights (QAMOO's actual method~\cite{qamoo}); and
(iii) \emph{BW-QAMOO} (Algorithm~\ref{alg:bandit}), with its GP-EHVI surrogate (seeded from the $m$
corners, budget cap $B{=}40$). Both BW-QAMOO and QAMOO (random) operate
grid-free, so their comparison is exact; QAMOO (uniform grid) is included as the naive baseline.
As evaluation metric, we use \emph{runs-to-target}: the number of scalarized QAOA solves a
strategy needs to reach $90\%$ and $95\%$ of the \emph{exact} Pareto hypervolume. Fewer runs to the
same target means a more query-efficient strategy. Every experiment averages over $10$ random seeds
(instances) per cell.

\section{Results}
\label{sec:results}

\begin{figure}[tb]
\centering
\includegraphics[width=\textwidth]{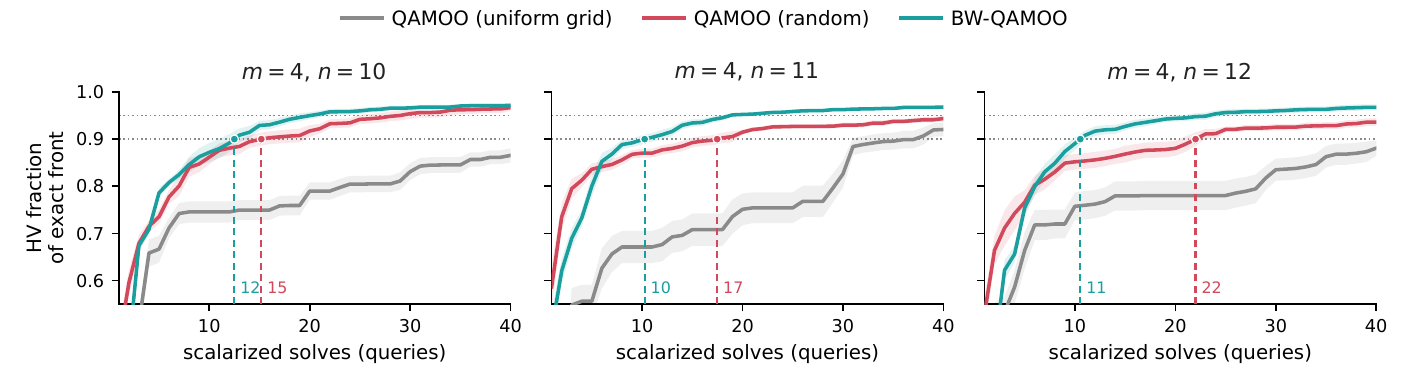}
\caption{Hypervolume reached versus number of scalarized solves on the simulated QAOA solver,
for $m{=}4$ objectives (mean over $10$ seeds, $\pm1$ standard-error band). Each curve is the fraction
of the exact Pareto hypervolume secured after $1,2,3,\dots$ queries, for QAMOO (uniform grid) (grey),
QAMOO (random) (red), and BW-QAMOO (teal); dotted horizontal lines mark the $90/95\%$ targets.
The vertical dashed lines drop from where the \emph{mean} BW-QAMOO (teal) and QAMOO (random) (red)
curves first reach the $90\%$ target, with that crossing query annotated on the axis; their separation
is the query saving. Because these crossings are read off the averaged curve, they differ slightly from
the per-instance mean runs-to-target in Table~\ref{tab:runs} (which averages each instance's own
crossing). Up-and-to-the-left is better: BW-QAMOO crosses $90\%$ well left of QAMOO (random) (e.g.\ $10$
vs.\ $17$ queries at $n{=}11$) with a tighter band, while QAMOO (uniform grid) trails. ($m{=}3$ shows the
same ordering, see Table~\ref{tab:runs}.)}
\label{fig:conv-qaoa}
\end{figure}

\begin{table}[tb]
\centering
\small
\caption{QAOA queries to reach a target hypervolume fraction (runs-to-target, mean over $10$ seeds;
fewer is better). Columns are BW-QAMOO (\textbf{BW}), QAMOO (random) (\textbf{R}), and QAMOO (uniform
grid) (\textbf{G}). The ``sig.'' row gives one-sided paired Wilcoxon tests that BW-QAMOO needs fewer solves than the
baseline in that column (R or G), pooling $30$ instances per $(m,\text{target})$
(\textbf{***}\,$p{<}10^{-3}$, \textbf{**}\,$p{<}10^{-2}$, \textbf{*}\,$p{<}0.05$).}
\label{tab:runs}
\begin{tabular}{c ccc ccc @{\hskip 1.4em} ccc ccc}
\toprule
& \multicolumn{6}{c}{$m{=}3$} & \multicolumn{6}{c}{$m{=}4$} \\
\cmidrule(lr){2-7}\cmidrule(lr){8-13}
& \multicolumn{3}{c}{\emph{90\%}} & \multicolumn{3}{c}{\emph{95\%}}
& \multicolumn{3}{c}{\emph{90\%}} & \multicolumn{3}{c}{\emph{95\%}} \\
\cmidrule(lr){2-4}\cmidrule(lr){5-7}\cmidrule(lr){8-10}\cmidrule(lr){11-13}
$n$ & BW & R & G & BW & R & G & BW & R & G & BW & R & G \\
\midrule
10 & \textbf{8.5} & 10.7 & 17.2 & \textbf{14.8} & 25.7 & 21.7 & \textbf{12.4} & 17.5 & 46.2 & \textbf{21.2} & 26.2 & 58.8 \\
11 & \textbf{5.0} & 6.2 & 12.5 & \textbf{8.7} & 23.8 & 16.2 & \textbf{11.2} & 15.6 & 35.5 & \textbf{19.1} & 37.0 & 50.8 \\
12 & \textbf{8.2} & 12.9 & 16.5 & \textbf{19.2} & 24.7 & 32.3 & \textbf{11.8} & 19.2 & 44.7 & \textbf{22.9} & 37.5 & 59.1 \\
\cmidrule(l){1-13}
\emph{sig.} &  & \textbf{**} & \textbf{***} &  & \textbf{***} & \textbf{***}
            &  & \textbf{***} & \textbf{***} &  & \textbf{***} & \textbf{***} \\
\bottomrule
\end{tabular}
\end{table}

Figure~\ref{fig:conv-qaoa} and Table~\ref{tab:runs} are the central result. Against QAMOO's own
random-sampling method, BW-QAMOO reaches the same hypervolume with a \emph{median $33\%$} fewer
scalarized QAOA solves at the $90/95\%$ targets, winning \emph{all $12/12$} cells, and the saving is
significant by a paired Wilcoxon test (Table~\ref{tab:runs}). The advantage grows with $m$: the median saving rises from $29\%$ at $m{=}3$
to $34\%$ at $m{=}4$ (range $19$--$48\%$ across the $m{=}4$ cells), where the weight simplex is
three-dimensional and random sampling wastes more of its budget --- the teal curve crossing each target
line well left of the red one in the bottom row, with a far tighter standard-error band (BW-QAMOO is not
just faster but more reliable). Against
QAMOO (uniform grid) the gap is larger still ($61$--$74\%$ at $m{=}4$), because a fixed-resolution grid grows
combinatorially --- $84$ weights at $m{=}4$ versus $28$ at $m{=}3$ --- and spends most redundantly, the
``gridding is wasteful'' failure made concrete. Every run is one expensive query, so these are
proportional cost reductions. The QAOA result here is a \emph{simulation}, not a hardware run (see
limitations).

\section{Discussion, limitations, and open challenges}
\label{sec:disc}

The saving is consistent and, in this setting, sizable, and it grows precisely where it matters: as
the objective count rises and the weight simplex expands, blind sampling wastes an increasing share of
the budget while an adaptive selector concentrates queries on the weights that still enlarge the
front. Since each query is one expensive quantum solve, these are direct, proportional reductions in
the cost of tracing a front on a quantum device.

We read this as evidence for a broader and, we think, promising direction: the mature machinery of
classical machine learning and BO can substantially improve how near-term quantum
optimizers are used, without any change to the quantum subroutine itself --- the solver stays an
expensive oracle, and a classical surrogate decides what to ask it. The same idea should transfer to
other quantum-optimization pipelines that sweep a family of related problems.

This is a deliberately narrow study: a frustrated-Ising benchmark at $n\le12$ with random couplings,
kept small so the exact Pareto front is enumerable for scoring, and --- most importantly --- run on a
noiseless QAOA \emph{simulation} rather than real hardware. The decisive next step is a deployment on an actual quantum
device, where gate noise, readout error, and queue latency all bear on the per-query cost the method
sets out to reduce; there, the GP's calibrated variance could do double duty, steering queries toward
the weights a noisy device solves reliably. Beyond that, a proxy reference point would push the
measurement past the enumerable regime to larger $n$ and $m$; a scalarization that combines the
objectives non-linearly, rather than by the weighted sum we use, would reach the non-convex parts of
the front a weighted sum provably cannot; a policy meta-learned across
instances could carry savings between related alloy compositions; fitted cluster interactions in
place of random couplings~\cite{crystalconfig,heaanneal} would close the remaining gap to a real
materials-design workflow; and extending the site variable from two species to the
\emph{multi-component} alloys (ternaries, high-entropy alloys) where the field's interest concentrates
would broaden the applicability --- a change to how each objective is encoded into qubits, not to the
weight-selection loop, which is agnostic to the encoding beneath it (Appendix~\ref{app:multispecies}).

\appendix

\section{Background: the Ising model and its link to our problem}
\label{app:ising}

The whole paper --- the benchmark, the scalarization, and the QAOA solver --- rests on the
\emph{Ising model}, so we state it here for readers coming from the materials or machine-learning
side rather than physics.

\paragraph{The model.} An Ising model is an energy defined over $n$ binary variables, the
\emph{spins} $\sigma_i\in\{-1,+1\}$. Its energy is a sum of \emph{on-site fields} and
\emph{pairwise couplings},
\begin{equation}
E(\bs)=\sum_i h_i\,\sigma_i+\sum_{i<j}J_{ij}\,\sigma_i\sigma_j,
\qquad \bs=(\sigma_1,\dots,\sigma_n)\in\{-1,+1\}^n,
\label{eq:ising-app}
\end{equation}
where the field $h_i$ biases spin $i$ toward $+1$ or $-1$, and the coupling $J_{ij}$ rewards or
penalizes a pair for being aligned ($\sigma_i\sigma_j=+1$) or anti-aligned ($\sigma_i\sigma_j=-1$).
Minimizing $E$ over the $2^n$ spin configurations is the combinatorial problem of interest; it is
NP-hard in general~\cite{lucas}, and no more than a quadratic (degree-two) polynomial in the spins is
needed to encode a remarkable range of hard problems~\cite{lucas}. A change of variable
$\sigma_i=1-2x_i$ turns Eq.~\eqref{eq:ising-app} into an equivalent \emph{QUBO} (quadratic
unconstrained binary optimization) over bits $x_i\in\{0,1\}$; the two forms are interchangeable and
we use whichever is convenient.

\paragraph{Frustration --- what makes it hard.} If all couplings could be satisfied at once the
ground state would be trivial. \emph{Frustration} is the situation where they cannot: on a loop of
spins with competing coupling signs, no configuration satisfies every bond simultaneously, so the
energy landscape develops many competing local minima. This is precisely the regime our benchmark
targets (random signed couplings; Appendix~\ref{app:objectives}), and it is what makes the
configurational ground state non-trivial to find.

\paragraph{Why it fits our problem.} In an alloy or crystal, the design variable is which of two
species occupies each of $n$ lattice sites --- a natural binary spin $\sigma_i\in\{-1,+1\}$. To
leading order in a \emph{cluster expansion} (the standard way configurational properties of alloys
are represented~\cite{sanchez,ceder,crystalconfig}), each configurational property is a constant plus
on-site terms plus pairwise terms --- exactly the Ising form of Eq.~\eqref{eq:ising-app}, with
$h_i,J_{ij}$ the fitted effective interactions. So a materials-configuration objective \emph{is} an Ising energy;
Appendix~\ref{app:objectives} makes this correspondence explicit for each objective we use.

\paragraph{Why it fits quantum optimizers.} Each spin $\sigma_i$ maps to a qubit through the Pauli-$Z$
operator, and Eq.~\eqref{eq:ising-app} becomes a diagonal \emph{cost Hamiltonian}
$H_C=\sum_i h_i Z_i+\sum_{i<j}J_{ij}Z_iZ_j$ whose ground state is the minimizing configuration. Both
quantum annealers~\cite{kadowaki} and gate-model methods such as QAOA (Appendix~\ref{app:qamoo}) are
built to seek that ground state, which is why the Ising encoding makes materials-configuration search
a direct target for quantum optimization. A single weighted sum of several Ising energies,
$E_{\bw}=\sum_k w_k E_k$ (Eq.~\eqref{eq:ising}), is \emph{again} of the form
Eq.~\eqref{eq:ising-app} --- the property BW-QAMOO relies on, since every scalarized weight is then
itself a valid Ising (hence QAOA) instance.

\section{Background: QAOA and QAMOO}
\label{app:qamoo}

\paragraph{QAOA.} The Quantum Approximate Optimization Algorithm~\cite{qaoa} is a variational
gate-model method for minimizing an Ising cost such as Eq.~\eqref{eq:ising}. The cost is written as a
diagonal \emph{cost Hamiltonian} $H_C=\sum_i h_i Z_i+\sum_{i<j}J_{ij}Z_iZ_j$ (each spin $\sigma_i$
becomes a qubit Pauli-$Z_i$), and a \emph{mixer} $H_M=\sum_i X_i$ drives transitions between
bitstrings. Starting from the uniform superposition $\ket{+}^{\otimes n}$, QAOA applies $p$
alternating layers,
\begin{equation}
\ket{\boldsymbol{\gamma},\boldsymbol{\beta}}
=\prod_{\ell=1}^{p} e^{-i\beta_\ell H_M}\,e^{-i\gamma_\ell H_C}\ \ket{+}^{\otimes n},
\end{equation}
parameterized by $2p$ angles $(\boldsymbol{\gamma},\boldsymbol{\beta})$. A classical optimizer
(here COBYLA) tunes the angles to minimize the measured expected energy
$\langle H_C\rangle$; the final state is sampled ($4096$ shots) and the lowest-energy
bitstring observed is returned as the solution. Depth $p$ trades circuit expressivity against noise;
we use $p{=}2$. Each such solve --- angle optimization plus sampling --- is one ``expensive query'' in
the main text. We run QAOA on the Qiskit Aer simulator with shot-based readout and \emph{no} noise
model: it confirms the algorithm behaves as intended but omits the device noise, readout error, and
queueing of real hardware.

\begin{figure}[t]
\centering
\resizebox{\textwidth}{!}{%
\begin{tikzpicture}[
    wire/.style={thick},
    gate/.style={draw, fill=white, thick, minimum height=7mm, minimum width=8mm, rounded corners=1pt},
    zz/.style={draw, fill=blue!8, thick, minimum height=7mm, minimum width=9mm, rounded corners=1pt},
    dot/.style={circle, fill=black, inner sep=1.6pt},
    x=1cm, y=1cm,
]
\foreach \q in {0,1,2} {
  \draw[wire] (-0.4,-\q) -- (14.2,-\q);
  \node[left] at (-0.5,-\q) {$\ket{0}_{\q}$};
}
\foreach \q in {0,1,2} { \node[gate] (H\q) at (0.6,-\q) {$H$}; }
\node[above=5mm] at (0.6,0) {\footnotesize init $\ket{+}^{\otimes n}$};

\foreach \q in {0,1,2} { \node[gate] (RZa\q) at (2.4,-\q) {$R_Z$}; }
\node[zz] (ZZa01) at (4.3,-0.5) {$R_{ZZ}$};
\draw[thick] (ZZa01.north) -- (4.3,0); \node[dot] at (4.3,0){};
\draw[thick] (ZZa01.south) -- (4.3,-1); \node[dot] at (4.3,-1){};
\node[zz] (ZZa12) at (5.7,-1.5) {$R_{ZZ}$};
\draw[thick] (ZZa12.north) -- (5.7,-1); \node[dot] at (5.7,-1){};
\draw[thick] (ZZa12.south) -- (5.7,-2); \node[dot] at (5.7,-2){};
\foreach \q in {0,1,2} { \node[gate] (RXa\q) at (7.0,-\q) {$R_X$}; }

\foreach \q in {0,1,2} { \node[gate] (RZb\q) at (8.6,-\q) {$R_Z$}; }
\node[zz] (ZZb01) at (10.0,-0.5) {$R_{ZZ}$};
\draw[thick] (ZZb01.north) -- (10.0,0); \node[dot] at (10.0,0){};
\draw[thick] (ZZb01.south) -- (10.0,-1); \node[dot] at (10.0,-1){};
\node[zz] (ZZb12) at (11.2,-1.5) {$R_{ZZ}$};
\draw[thick] (ZZb12.north) -- (11.2,-1); \node[dot] at (11.2,-1){};
\draw[thick] (ZZb12.south) -- (11.2,-2); \node[dot] at (11.2,-2){};
\foreach \q in {0,1,2} { \node[gate] (RXb\q) at (12.4,-\q) {$R_X$}; }
\foreach \q in {0,1,2} { \node[gate, fill=gray!12] (M\q) at (13.7,-\q) {\footnotesize $\displaystyle\measuredangle$}; }

\begin{scope}[on background layer]
  \node[fit=(RZa0)(ZZa01)(ZZa12)(RXa2), draw=orange!70, thick, dashed, rounded corners,
        inner sep=3.5mm, label={[orange!80!black]above:{\footnotesize layer 1 \; $e^{-i\beta_1 H_M}e^{-i\gamma_1 H_C}$}}] {};
  \node[fit=(RZb0)(ZZb01)(ZZb12)(RXb2), draw=blue!60, thick, dashed, rounded corners,
        inner sep=3.5mm, label={[blue!70!black]above:{\footnotesize layer 2 \; $e^{-i\beta_2 H_M}e^{-i\gamma_2 H_C}$}}] {};
\end{scope}
\node[align=left, anchor=west] at (-0.4,-3.05)
  {\footnotesize $R_Z(\gamma_\ell\,c_i)$: single-qubit $Z_i$ field term \quad
   $R_{ZZ}(\gamma_\ell b_{ij})$: coupling $Z_iZ_j$ (a \textsc{cx}--$R_Z$--\textsc{cx} gadget) \quad
   $R_X(2\beta_\ell)$: mixer \quad $\measuredangle$: computational-basis read-out};
\end{tikzpicture}%
}
\caption{The QAOA ansatz actually run (depth $p{=}2$), on a representative $3$-qubit slice with two
coupled pairs. After a Hadamard layer prepares $\ket{+}^{\otimes n}$, each of the $p$ layers applies
the cost unitary $e^{-i\gamma_\ell H_C}$ --- an $R_Z$ per single-$Z$ term and an $R_{ZZ}$ per
coupled pair of Eq.~\eqref{eq:pauli-cost} --- followed by the mixer $e^{-i\beta_\ell H_M}$, one $R_X$
per qubit. The $2p{=}4$ angles $(\gamma_1,\beta_1,\gamma_2,\beta_2)$ are tuned by COBYLA; the final
state is measured at $4096$ shots.}
\label{fig:qaoa-circuit}
\end{figure}
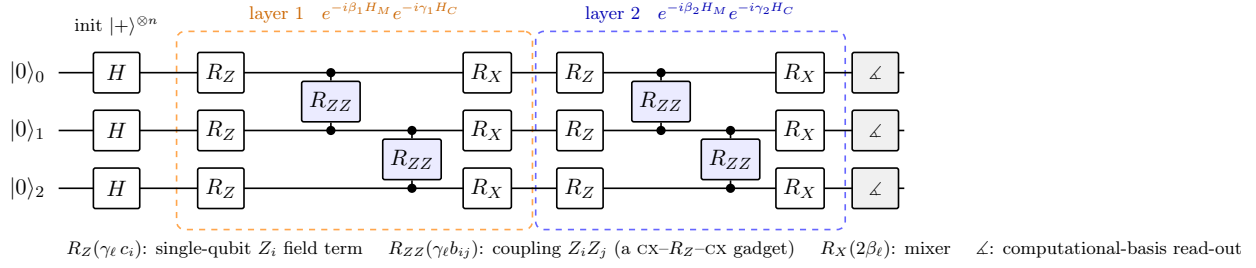

\paragraph{Ansatz, gates, and the Pauli cost Hamiltonian (exact circuit run).}
We use the standard QAOA ansatz (Qiskit's \texttt{QAOAAnsatz}) with $p{=}2$, so the circuit carries
$2p{=}4$ variational parameters $(\gamma_1,\beta_1,\gamma_2,\beta_2)$. Each objective is scalarized
to an Ising energy, converted to an equivalent QUBO in binary variables $x_i\in\{0,1\}$ (via
$\sigma_i{=}1{-}2x_i$), and the QUBO is re-expressed as a diagonal cost Hamiltonian using
$x_i=\tfrac{1}{2}(1-Z_i)$. For a QUBO $\sum_i a_i x_i+\sum_{i<j}b_{ij}x_ix_j$ this yields
\begin{equation}
H_C=\underbrace{\Big(\tfrac12\textstyle\sum_i a_i+\tfrac14\sum_{i<j}b_{ij}\Big)}_{\text{constant offset}}\,I
-\sum_i\Big(\tfrac12 a_i+\tfrac14\textstyle\sum_{j\neq i}b_{ij}\Big)Z_i
+\sum_{i<j}\tfrac14 b_{ij}\,Z_iZ_j,
\label{eq:pauli-cost}
\end{equation}
i.e.\ single-qubit $Z_i$ terms (a local field) and two-qubit $Z_iZ_j$ terms on every coupled pair,
plus an identity offset that shifts the energy but not the minimizer. Concretely the cost layer
$e^{-i\gamma_\ell H_C}$ compiles to an $R_{Z}(\theta)=e^{-i\theta Z/2}$ rotation for each $Z_i$ term
and an $R_{ZZ}(\theta)=e^{-i\theta Z_iZ_j/2}$ interaction (a \textsc{cx}--$R_Z$--\textsc{cx} gadget)
for each $Z_iZ_j$ term; the mixer layer $e^{-i\beta_\ell H_M}$ with $H_M=\sum_i X_i$ compiles to an
$R_X(2\beta_\ell)$ on every qubit. The initial state $\ket{+}^{\otimes n}$ is prepared by a layer of
Hadamards. During angle optimization the expectation $\langle H_C\rangle$ is evaluated on the Aer
\texttt{EstimatorV2} primitive; COBYLA (max.\ $100$ iterations) tunes the angles from a seeded random
start $\gamma,\beta\sim\mathcal{U}(0,\pi)$; the tuned circuit is then measured with the
\texttt{SamplerV2} primitive at $4096$ shots, and among the sampled bitstrings we return the one of
lowest QUBO energy. This ``tuned'' readout isolates the query-efficiency question from sampler noise;
every scalarized weight is one such circuit solve. Figure~\ref{fig:qaoa-circuit} shows the resulting
gate-level circuit on a representative $3$-qubit slice.

\paragraph{QAMOO.} Quantum Approximate Multi-Objective Optimization~\cite{qamoo} extends QAOA from a
single objective to a Pareto front over $m$ objectives. Its mechanism is \emph{scalarization}: for a
weight vector $\bw$ on the simplex it forms the single Ising cost $E_\bw$ of
Eq.~\eqref{eq:ising}, solves that one problem with QAOA, and records the returned
configuration as one candidate Pareto point. Sweeping $\bw$ across the simplex and keeping the
non-dominated configurations traces an approximation to the front. Crucially, QAMOO chooses the
weights by \emph{random sampling}: rather than discretizing the simplex on a grid, it samples $\bw$
uniformly at random from the feasible set, ``without having to enumerate an exponential number of
grid points''~\cite{qamoo}. Because every $E_\bw$ is itself a valid Ising problem, no change to the
quantum subroutine is needed --- only a choice of which weights to solve. This paper replaces the
random draw with an adaptive, acquisition-driven selector (Algorithm~\ref{alg:bandit}) while leaving
the per-weight QAOA solve untouched.

\section{The GP-EHVI surrogate in detail}
\label{app:gp}

\paragraph{The GP-EHVI surrogate.} We fit one Gaussian-process regressor~\cite{rw06} per objective
(Mat\'ern-$5/2$ kernel, an anisotropic length scale per simplex coordinate), giving a posterior
\emph{mean and variance} for each unqueried weight rather than a point estimate. The acquisition is an
\emph{optimistic} (upper-confidence-bound) surrogate for the expected hypervolume improvement in
Eq.~\eqref{eq:ehvi}: for each candidate weight it forms the optimistic objective vector
$\hat{\bs}(\bw)=\boldsymbol{\mu}(\bw)-\kappa\,\boldsymbol{\sigma}(\bw)$ ($\kappa{=}1.5$; the objectives
are minimized, so subtracting the posterior standard deviation is the optimistic direction) and scores
the hypervolume gain that this single point would contribute,
$\mathrm{HV}(\mathcal{P}_t\cup\{\hat{\bs}(\bw)\})-\mathrm{HV}(\mathcal{P}_t)$ --- one deterministic
hypervolume evaluation per candidate, no Monte-Carlo sampling. The $\kappa\boldsymbol{\sigma}$ term now
supplies genuine exploration, so the selector can prefer a weight whose outcome is \emph{uncertain} over
one whose predicted mean gain is merely high --- the explore/exploit balance at the heart of Bayesian
optimization~\cite{shahriari}. This is a cheap upper-confidence-bound realization of the GP-EHVI
acquisition of multi-objective Bayesian optimization~\cite{ehvi,qehvi}, ported to the weight simplex; a
full Monte-Carlo EHVI (averaging the gain over posterior draws from the same surrogate) is a drop-in
alternative at higher per-step cost.

\paragraph{Beyond this benchmark.} The GP's advantages should widen further where this benchmark's
assumptions break: on \emph{noisy or stochastic solvers} (real hardware), where its calibrated
variance can route queries to weights the device solves reliably; under \emph{non-smooth or
multi-modal} weight-to-objective maps (e.g.\ across a ground-state phase transition); and under
\emph{non-linear scalarizations} --- which combine the objectives by something other than a weighted
sum, so as to reach non-convex parts of the front, at the price of a less smooth weight-to-objective
map that the surrogate must then learn. These are left to future work (Sec.~\ref{sec:disc}).

\section{From crystal-design objectives to Ising energies}
\label{app:objectives}

Our benchmark is a \emph{synthetic} instantiation of crystal/alloy design, and this appendix makes the
correspondence explicit and states plainly where it is idealized. The design variable is which of two
species occupies each of the $n$ symmetry-distinct sites of a fixed lattice, encoded as a spin
$\sigma_i\in\{-1,+1\}$ (species A/B). Every objective below is a function of the occupation pattern
$\bs$, and each is naturally \emph{quadratic} in the spins --- a constant, plus on-site fields
$h_i\sigma_i$, plus pairwise couplings $J_{ij}\sigma_i\sigma_j$ --- which is exactly the Ising form
Eq.~\eqref{eq:ising} that QAOA minimizes. This quadratic-in-occupation structure is not incidental: it
is the leading order of a \emph{cluster expansion}~\cite{sanchez,ceder}, the standard way
configurational properties of alloys are represented, so the mapping to Ising is the physically
motivated one.

\paragraph{Candidate objectives and their Ising terms.}
\begin{itemize}
\item \textbf{Configurational / formation energy} ($E_1$ here). The effective interactions between
occupied sites: $h_i$ a site preference, $J_{ij}$ a pair interaction (negative $J_{ij}$ favors
like neighbors, i.e.\ clustering/phase separation; positive favors unlike neighbors, i.e.\ ordering).
Frustration --- competing signs on a loop that cannot all be satisfied --- is exactly what makes the
configurational ground state non-trivial, which is why $E_1$ uses random signed couplings.
\item \textbf{Elastic / strain-mismatch energy.} Size mismatch between the two species stores elastic
energy that depends on which neighbors are unlike; to leading order this is again $\sum_{ij}
K^{\text{el}}_{ij}\sigma_i\sigma_j$ with $K^{\text{el}}_{ij}$ set by the bond stiffness and the atomic
size difference. Minimizing it favors arrangements that relieve strain.
\item \textbf{Chemical short-range order (disorder).} A target degree of mixing is written as a
penalty on the count of like-neighbor pairs, $\sum_{ij}\!J^{\text{sro}}_{ij}\sigma_i\sigma_j$ (up to a
constant); tuning its sign drives the configuration toward ordering or toward a solid solution.
\item \textbf{Cost / scarcity.} The price or supply risk of the chosen composition is
\emph{linear} in occupancy, $\sum_i c_i\sigma_i$ (a pure field term $h_i=c_i/2$, no couplings): a
cheaper species on more sites lowers it. Density and other composition-linear properties enter the
same way.
\end{itemize}
Because each objective is a weighted sum of fields and couplings, a scalarization $\sum_k w_k E_k$ is
\emph{itself} an Ising energy (its fields and couplings are the $\bw$-weighted sums; Eq.~\eqref{eq:ising}),
which is what lets a single QAOA solve handle any weight $\bw$.

\paragraph{What the synthetic model does and does not capture.} Our $E_2,\dots,E_m$ use fixed, distinct
\emph{random} coupling matrices rather than the $h_i,J_{ij}$ one would fit from DFT or a measured
cluster expansion, so the numbers here reflect the \emph{structure} of the problem --- several
competing quadratic objectives over a frustrated configuration space, with a weight simplex whose
dimension grows with $m$ --- but not any specific material. The query-efficiency result depends only
on that structure (a family of related scalarized Ising solves swept over $\bw$), not on the couplings
being physical, so it should carry over once real interactions are substituted; conversely, we make no
claim that the resulting fronts correspond to real materials trade-offs. Replacing the random
couplings with fitted cluster interactions --- and moving beyond two species on a single sublattice
(Appendix~\ref{app:multispecies}) --- is the natural next step (Sec.~\ref{sec:disc}).

\section{More than two species per site}
\label{app:multispecies}

A binary spin $\sigma_i\in\{-1,+1\}$ names exactly two species, so the encoding of
Appendix~\ref{app:objectives} models a \emph{binary} alloy. Multi-component alloys --- ternaries and,
in the extreme, high-entropy alloys~\cite{heaanneal} --- put $q>2$ species on each site, and a single
binary variable per site can no longer name the choice. We stress at the outset that this is a change
to the \emph{per-objective encoding} only: BW-QAMOO treats each scalarized solve as a black box and is
indifferent to how many qubits and terms that box contains, so its weight-selection loop, and the
query-efficiency result, are unaffected. What changes is the qubit count and structure of each solve.
There are three standard routes, each with a concrete cost.

\paragraph{One-hot encoding.} Give each site $q$ binary variables $x_{i,k}\in\{0,1\}$, one per species,
and enforce ``exactly one species per site'' with a penalty $\lambda\big(\sum_k x_{i,k}-1\big)^2$. The
penalty is quadratic in the bits, so the problem \emph{remains} an Ising/QUBO instance solvable by the
same QAOA subroutine~\cite{lucas}; the price is $n\,q$ qubits rather than $n$ and a penalty weight
$\lambda$ that must be tuned (too small admits infeasible one-hot patterns, too large drowns the
objective). This is the most common choice and the one that most directly preserves our pipeline.

\paragraph{Binary / domain-wall encoding.} Name the $q$ species with $\lceil\log_2 q\rceil$ bits per
site, using exponentially fewer qubits. The objective in these bits generally acquires terms of degree
$>2$ that must then be quadratized (via ancillas) to return to Ising form; domain-wall
encodings~\cite{chancellor} keep the required couplings local and are a practical middle ground between
qubit count and Hamiltonian complexity.

\paragraph{Potts / higher-spin objectives.} Physically, $q>2$ discrete states per site is a
\emph{Potts} model rather than an Ising one, the natural generalization for multi-component ordering;
to run it on gate-model QAOA it is reduced to qubits by one of the two routes above. In every case the
scalarization $\sum_k w_k E_k$ stays a single problem of the same (encoded) form, so QAMOO's sweep ---
and our adaptive selection of which weights to sweep --- applies unchanged.

\end{document}